\documentclass[11pt]{article}
\usepackage[margin=0.88in]{geometry}
\usepackage{amsmath,amssymb,amsthm,mathtools,stmaryrd}
\usepackage{microtype}
\usepackage{booktabs,array,tabularx}
\usepackage{enumitem}
\usepackage{tikz}
\usetikzlibrary{arrows.meta,positioning,calc,fit,backgrounds,matrix,decorations.pathreplacing}
\usepackage{xcolor}
\definecolor{linkblue}{RGB}{24,76,130}
\definecolor{theoremblue}{RGB}{238,246,252}
\definecolor{figblue}{RGB}{45,96,150}
\definecolor{figgreen}{RGB}{54,117,94}
\definecolor{figred}{RGB}{158,63,66}
\definecolor{figgray}{RGB}{92,101,111}
\definecolor{softgray}{RGB}{246,247,249}
\usepackage[colorlinks=true,linkcolor=linkblue,citecolor=linkblue,urlcolor=linkblue]{hyperref}
\usepackage[nameinlink,capitalize]{cleveref}
\usepackage{mathpazo}
\usepackage[T1]{fontenc}
\usepackage{authblk}
\usepackage{caption}
\setlist{nosep,leftmargin=*}
\newtheorem{theorem}{Theorem}[section]
\newtheorem{lemma}[theorem]{Lemma}
\newtheorem{proposition}[theorem]{Proposition}
\newtheorem{corollary}[theorem]{Corollary}
\newtheorem{definition}[theorem]{Definition}
\newtheorem{remark}[theorem]{Remark}

\newcommand{\R}{\mathbb R}
\newcommand{\C}{\mathbb C}

\newcommand{\cH}{\mathcal H}

\newcommand{\wt}{\operatorname{wt}}

\newcommand{\Enc}{\operatorname{Enc}}

\newcommand{\ket}[1]{\lvert #1\rangle}
\newcommand{\bra}[1]{\langle #1\rvert}
\newcommand{\braket}[2]{\langle #1\mid #2\rangle}

\title{\bfseries Distributed Phase Tags:\\A Locality-Preserving Realification Compiler \\ for Misty-State Quantum Programs}
\author[1]{Dan-Adrian German}
\author[2]{Tony Newton\thanks{Newton Astro Labs is the trading name under which Tony Newton conducts independent computational research in the United Kingdom; it is not a limited company.}}
\affil[1]{Indiana University Computer Science, Bloomington, IN, USA\\\texttt{dgerman@iu.edu}}
\affil[2]{Newton Astro Labs*, London, UK\\\texttt{tony.newton79@gmail.com}}
\date{} 

\begin{document}
\maketitle

\begin{abstract}
Complex amplitudes can be simulated over real Hilbert spaces by adjoining a two-dimensional real/imaginary degree of freedom.  In a multipartite program, however, one globally shared phase tag is not a locality-preserving program transformation: every party would have to act on the same auxiliary system.  We recast the distributed real simulation of multipartite quantum mechanics as a syntax-directed compiler into a real misty-state intermediate language.  For $n$ parties we give explicit phase-reference codewords $\ket{0}_{P_n}$ and $\ket{1}_{P_n}$ on $n$ local tags and prove that the same real operator $J=XZ$ on \emph{any} single tag realizes the logical action of multiplication by $i$.  If a complex party-local gate is $A=A_R+iA_I$, its real translation is
\[
   \widehat A_j=A_R\otimes I+A_I\otimes J_j,
\]
which acts only on party $j$'s data and local phase tag.  We prove an intertwining theorem for arbitrary finite party-local pure-state circuits, exact preservation of computational-basis measurement probabilities, preservation of the run-time interaction graph, and one-rebit-per-party overhead.  For the canonical three-party GHZ program, the translation lies entirely inside the ordinary black-and-white misty gate fragment: controlled-$J$ replaces the complex phase operation, while the invariant $J_jJ_k\ket{0}_P=-\ket{0}_P$ transfers the required branch sign.  This yields a short correctness proof for all four promised GHZ inputs and explains the earlier 64-path cancellation calculation at a single logical phase-code boundary.  We distinguish this \emph{program realification} problem from the weaker task of finding any real strategy with the same correlations, and state precisely which parts are known real-Hilbert-space simulation and which parts belong to the misty-language compilation layer developed here.
\end{abstract}

\section{Introduction and main results}\label{sec:intro}

Quantum programs are normally written over complex Hilbert spaces.  A familiar realification replaces a complex vector by its real and imaginary components, encoded in an additional two-level real system.  Rudolph and Grover used this idea in the circuit model to show that universal quantum computation can be carried out with real amplitudes \cite{RudolphGrover2002}; McKague, Mosca and Gisin subsequently gave a multipartite construction that preserves local evolution and measurements by distributing the real/imaginary degree of freedom among the parties \cite{McKague2009}.

The programming-language issue is more specific.  A translation of a distributed program should preserve more than its final input--output statistics.  If the source program consists of party-local commands, a target program that makes every party access one globally shared phase ancilla has changed the ownership structure of the program.  For a nonlocality protocol such as GHZ, that distinction is especially visible: pre-shared entanglement is allowed, but run-time communication between the separated parties is not.

The misty-state formalism provides a concrete target syntax for asking this question.  Rudolph's black-and-white ``mist'' represents signed superpositions by collections of tokens and realizes interference by replacement and cancellation \cite{Rudolph2017}.  German subsequently formulated a term-rewriting semantics for pure quantum states \cite{German2026}, and Newton--German separated raw amplitude addition from physical normalization and supplied an exact amplitude semantics for finite mist expressions \cite{NewtonGerman2026}.  Rudolph, Newton, and German gave an explicit six-rebit misty-state construction showing that three parties can implement the canonical GHZ \(X/Y\) strategy using only real-valued operations \cite{ghzbw}. The present paper reorganizes and strengthens that construction as a compiler theorem.

\paragraph{The central distinction.}
There are two different tasks:
\begin{enumerate}[label=(\roman*)]
\item \emph{strategy realification}: find any real state and observables reproducing a target correlation;
\item \emph{program realification}: translate a specified complex state, specified local commands, and specified party decomposition into real syntax while preserving their compositional and locality structure.
\end{enumerate}
The first can be strictly easier.  For the GHZ correlation there is a three-rebit graph-state strategy; the six-rebit construction is therefore not a dimension lower bound for the game.  Its purpose is that it locally compiles the \emph{canonical complex GHZ program}.

\paragraph{Main theorem.}
For $n$ parties, introduce one phase tag per party and a two-dimensional logical code space $P_n=\operatorname{span}\{\ket{0}_{P_n},\ket{1}_{P_n}\}$.  The code is chosen so that the real matrix
\[
  J=XZ=\begin{pmatrix}0&-1\\1&0\end{pmatrix},\qquad J^2=-I,
\]
acts identically on the logical phase coordinate regardless of which party's tag it touches.  If $A=A_R+iA_I$ is a source operation owned by party $j$, compile it to
\begin{equation}\label{eq:compilerintro}
  \widehat A_j=A_R\otimes I+A_I\otimes J_j.
\end{equation}
We prove that this translation intertwines the complex and real denotations, and therefore extends compositionally to finite party-local circuits.

\begin{theorem}[Locality-preserving realification, informal form]\label{thm:intro}
Let $C=A_m\cdots A_1$ be a finite pure-state circuit in which each command $A_r$ is owned by a single party.  Encode the complex state by distributing one phase tag to each party and replace each command by \eqref{eq:compilerintro}.  Then the real target circuit has the same computational-basis output distribution as $C$, every translated command acts only on the same party's data together with that party's tag, and no run-time edge is added to the party-interaction graph.  If the source gates are unitary, the translated gates are real orthogonal gates.
\end{theorem}

The theorem's algebraic substrate is the established real-Hilbert-space simulation \cite{RudolphGrover2002,McKague2009}.  Our contribution is the explicit phase-code presentation, its syntax-directed formulation as a misty-program compiler, the ownership theorem, and the gate-level GHZ realization within the ordinary black-and-white fragment.

\begin{figure}[t]
\centering
\begin{tikzpicture}[node distance=8mm and 9mm, every node/.style={font=\small},
 box/.style={draw=black,rounded corners=2pt,fill=white,minimum height=9mm,align=center,inner sep=4pt},
 arr/.style={-{Latex[length=2.4mm]},thick,draw=figblue}]
\node[box] (src) {complex local program\\$A_j=A_{R,j}+iA_{I,j}$};
\node[box,right=of src] (comp) {phase-tag compiler\\$i\mapsto J_j$};
\node[box,right=of comp] (tgt) {real local program\\$\widehat A_j=A_{R,j}\otimes I+A_{I,j}\otimes J_j$};
\node[box,below=of comp,fill=theoremblue] (code) {distributed phase code\\$J_j\ket0_P=\ket1_P,\;J_j\ket1_P=-\ket0_P$};
\node[box,below=of src] (own) {source owner\\party $j$};
\node[box,below=of tgt] (own2) {target owner\\data $j$ + tag $j$};
\draw[arr] (src)--(comp);
\draw[arr] (comp)--(tgt);
\draw[arr] (code)--(comp);
\draw[arr] (own)-- node[left,font=\scriptsize]{ownership} (src);
\draw[arr] (tgt)-- node[right,font=\scriptsize]{preserved} (own2);
\end{tikzpicture}
\caption{Compiler architecture.  The nonlocal phase code is prepared once; every dynamic command remains local to its source owner.}
\label{fig:compiler}
\end{figure}
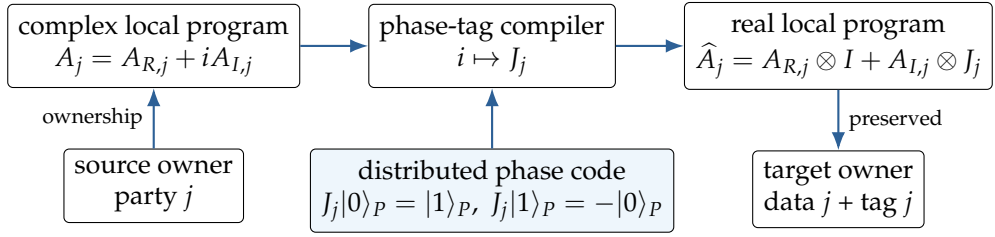

\paragraph{GHZ corollary.}
For three parties the logical phase-zero codeword is
\[
 \ket0_P=\frac12(\ket{000}-\ket{011}-\ket{101}-\ket{110}).
\]
The identity $J_jJ_k\ket0_P=-\ket0_P$ for $j\ne k$ is exactly the sign needed when two players execute the source $Y$ branch.  It turns the initial data state $(\ket{000}-\ket{111})/\sqrt2$ into $(\ket{000}+\ket{111})/\sqrt2$, after which three Hadamards expose the required even parity.  For the all-$X$ input no $J$ acts and the final parity is odd.  The long path expansion in the original derivation is therefore the concrete normal-form trace of one logical sign-transfer invariant.

\section{Misty syntax and real denotational semantics}\label{sec:syntax}

We use only a small fragment of the misty notation and make its denotation explicit.  Write $W$ and $B$ for white and black basis tokens,
\[
  \llbracket W\rrbracket=\ket0,\qquad \llbracket B\rrbracket=\ket1.
\]
For the finite fragment needed here,
\[
 t ::= W\mid B\mid -t\mid t_1\cdots t_r\mid\{t_1,\ldots,t_m\}\mid U(t).
\]
Juxtaposition is an ordered tensor product; braces collect amplitudes; the minus sign negates an amplitude; and gates extend linearly:
\begin{align*}
 \llbracket -t\rrbracket&=-\llbracket t\rrbracket,\\
 \llbracket t_1\cdots t_r\rrbracket&=\bigotimes_{j=1}^r\llbracket t_j\rrbracket,\\
 \llbracket\{t_1,\ldots,t_m\}\rrbracket&=\sum_{a=1}^m\llbracket t_a\rrbracket,\\
 \llbracket U(t)\rrbracket&=U\llbracket t\rrbracket.
\end{align*}
Thus a pair $t,-t$ cancels exactly.  Braces denote \emph{raw amplitude addition}; normalization is a separate operation.  This convention is essential: it prevents nested punctuation from being mistaken for physical normalization and agrees with the certified amplitude semantics of \cite{NewtonGerman2026}.

\subsection{The black-and-white gate fragment}
The Hadamard (PETE) action is
\[
  HW=\frac1{\sqrt2}\{W,B\},\qquad HB=\frac1{\sqrt2}\{W,-B\}.
\]
The real phase operator is
\begin{equation}\label{eq:J}
  J=XZ=\begin{pmatrix}0&-1\\1&0\end{pmatrix},
  \qquad JW=B,\quad JB=-W,\quad J^2=-I.
\end{equation}
Controlled-$J$ can be synthesized from controlled-$X$ and controlled-$Z$; controlled-$Z$ is obtained by conjugating CNOT by Hadamards on the target.  Hence the GHZ target program uses only real gates naturally expressible in the ordinary misty circuit language.

\subsection{A semantic real-linear intermediate language}
For the general compiler theorem we allow finite real matrices as target primitives.  This is a semantic intermediate language, not a claim that every real matrix is primitive in Rudolph's original gate vocabulary.  The distinction matters: the general theorem establishes a compositional semantics, while the GHZ case additionally admits a gate-level decomposition into the elementary black-and-white operations above.

A source command is a complex matrix $A$ acting on one party's local data space.  Relative to the computational basis write uniquely
\[
   A=A_R+iA_I,
   \qquad A_R,A_I\in M_d(\R).
\]
The target compiler will replace the scalar complex structure $i$ by the logical real operator encoded in the phase tags.

\section{The distributed phase-reference code}\label{sec:code}

The three-tag state used in the original GHZ derivation is the $n=3$ member of a simple family.  Let $x=(x_1,\ldots,x_n)\in\{0,1\}^n$, let $\wt(x)$ be its Hamming weight, and write $\ket{x}$ for the corresponding phase-tag basis vector.

\begin{definition}[Distributed phase code]\label{def:code}
For $n\ge1$, define
\begin{align}
 \ket0_{P_n}
 &=2^{-(n-1)/2}
   \sum_{\substack{x\in\{0,1\}^n\\ \wt(x)\;\mathrm{even}}}
   (-1)^{\wt(x)/2}\ket{x},\label{eq:p0n}\\
 \ket1_{P_n}
 &=2^{-(n-1)/2}
   \sum_{\substack{x\in\{0,1\}^n\\ \wt(x)\;\mathrm{odd}}}
   (-1)^{(\wt(x)-1)/2}\ket{x}.\label{eq:p1n}
\end{align}
\end{definition}

The supports are disjoint and each contains $2^{n-1}$ basis states, so the codewords are orthonormal.  They are invariant under permutation of the parties.

For $j\in\{1,\ldots,n\}$ let $J_j$ mean $J$ on phase tag $j$ and identity on every other tag.

\begin{lemma}[Local complex-structure lemma]\label{lem:Jlogical}
For every party $j$,
\begin{equation}\label{eq:Jlogical}
  J_j\ket0_{P_n}=\ket1_{P_n},\qquad
  J_j\ket1_{P_n}=-\ket0_{P_n}.
\end{equation}
Consequently the restriction of every $J_j$ to $P_n$ is the same logical matrix $J$, and for distinct $j,k$,
\begin{equation}\label{eq:twoJ}
   J_jJ_k\ket0_{P_n}=-\ket0_{P_n}.
\end{equation}
\end{lemma}

\begin{proof}
On a computational basis state, \eqref{eq:J} gives
\[
  J_j\ket{x}=(-1)^{x_j}\ket{x\oplus e_j}.
\]
Suppose first that $\wt(x)$ is even.  If $x_j=0$, flipping bit $j$ increases the weight by one and leaves the coefficient $(-1)^{\wt(x)/2}$ unchanged.  If $x_j=1$, the weight decreases by one and the extra factor $-1$ from $J$ is exactly the sign needed to obtain $(-1)^{(\wt(x\oplus e_j)-1)/2}$.  This proves $J_j\ket0_{P_n}=\ket1_{P_n}$.  The odd-weight calculation is the same with the signs reversed, giving $J_j\ket1_{P_n}=-\ket0_{P_n}$.  Applying the first identity and then the second yields \eqref{eq:twoJ}.
\end{proof}

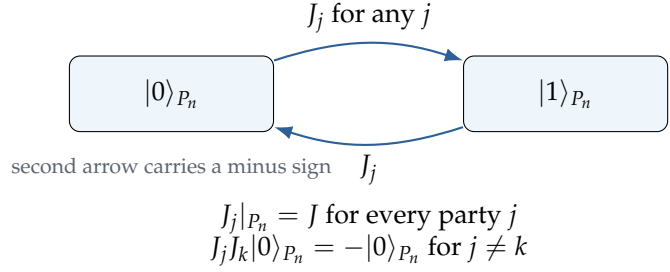
\begin{figure}[t]
\centering
\begin{tikzpicture}[node distance=25mm, every node/.style={font=\small},
 state/.style={draw=black,rounded corners=4pt,fill=theoremblue,minimum width=27mm,minimum height=10mm,align=center},
 arr/.style={-{Latex[length=2.4mm]},thick,draw=figblue}]
\node[state] (p0) {$\ket0_{P_n}$};
\node[state,right=of p0] (p1) {$\ket1_{P_n}$};
\draw[arr,bend left=18] (p0) to node[above] {$J_j$ for any $j$} (p1);
\draw[arr,bend left=18] (p1) to node[below] {$J_j$} (p0);
\node[below=2mm of p0,font=\scriptsize,text=figgray] {second arrow carries a minus sign};
\node[below=13mm of $(p0)!0.5!(p1)$,font=\small,align=center] {$J_j|_{P_n}=J$ for every party $j$\\$J_jJ_k\ket0_{P_n}=-\ket0_{P_n}$ for $j\neq k$};
\end{tikzpicture}
\caption{Logical complex structure from local real actions.  Every party has a different physical tag but the same logical action of multiplication by $i$.}
\label{fig:code}
\end{figure}

For $n=3$ the definition becomes
\begin{align}
 \ket0_P&=\frac12(\ket{000}-\ket{011}-\ket{101}-\ket{110}),\label{eq:p03}\\
 \ket1_P&=\frac12(\ket{001}+\ket{010}+\ket{100}-\ket{111}),\label{eq:p13}
\end{align}
which is exactly the phase-reference pair used in the compressed GHZ proof.

\section{A locality-preserving realification compiler}\label{sec:compiler}

We now state the general semantic translation.  Fix a computational basis of the source data Hilbert space $\cH\cong\C^N$.  Regard $\R^N$ as the corresponding real coordinate space.  Every source state has a unique decomposition
\[
   \ket\psi=\ket{\psi_R}+i\ket{\psi_I},\qquad
   \ket{\psi_R},\ket{\psi_I}\in\R^N.
\]

\begin{definition}[Distributed encoding]\label{def:encoding}
The $n$-party realification map is
\begin{equation}\label{eq:encoding}
 \Enc_n(\ket\psi)
 :=\ket{\psi_R}\otimes\ket0_{P_n}
   +\ket{\psi_I}\otimes\ket1_{P_n}.
\end{equation}
\end{definition}
Since the phase codewords are orthonormal,
\[
 \|\Enc_n(\psi)\|^2=\|\psi_R\|^2+\|\psi_I\|^2=\|\psi\|^2,
\]
so $\Enc_n$ is an isometry of real inner-product spaces.

Suppose a source command $A_j=A_{R,j}+iA_{I,j}$ is owned by party $j$.  The compiler emits
\begin{equation}\label{eq:Ahat}
  \widehat A_j
   :=A_{R,j}\otimes I_{p_j}+A_{I,j}\otimes J_{p_j},
\end{equation}
with identity on all other data and phase registers.

\begin{theorem}[Intertwining / compiler soundness]\label{thm:intertwine}
For every source state $\ket\psi$ and every party-local complex linear command $A_j$,
\begin{equation}\label{eq:intertwine}
   \widehat A_j\,\Enc_n(\ket\psi)
   =\Enc_n(A_j\ket\psi).
\end{equation}
If $A_j$ is unitary, then $\widehat A_j$ is a real orthogonal operator.
\end{theorem}

\begin{proof}
Using \cref{lem:Jlogical},
\begin{align*}
\widehat A_j\Enc_n(\psi)
&=(A_R\psi_R-A_I\psi_I)\otimes\ket0_{P_n}\\
&\quad +(A_R\psi_I+A_I\psi_R)\otimes\ket1_{P_n},
\end{align*}
which is exactly the real and imaginary decomposition of $A\psi$.  For the second statement, $A^\dagger A=I$ implies
\[
 A_R^TA_R+A_I^TA_I=I,
 \qquad A_R^TA_I=A_I^TA_R.
\]
Using $J^T=-J$ and $J^TJ=I$, direct multiplication gives $\widehat A_j^T\widehat A_j=I$.
\end{proof}

\begin{corollary}[Finite-circuit compositionality]\label{cor:circuit}
Let $C=A_m\cdots A_1$ be a finite sequence of party-local source commands and let $\widehat C=\widehat A_m\cdots\widehat A_1$.  Then
\[
   \widehat C\,\Enc_n(\psi)=\Enc_n(C\psi)
\]
for every source state $\psi$.
\end{corollary}

\begin{proof}
Induct on the command sequence using \cref{thm:intertwine}.
\end{proof}

\subsection{Measurement preservation}
Let $\Pi_x=\ket{x}\!\bra{x}$ project onto a computational-basis outcome of the data registers.  Since $\ket0_{P_n}$ and $\ket1_{P_n}$ are orthonormal,
\begin{equation}\label{eq:prob}
 \| (\Pi_x\otimes I_P)\Enc_n(\psi)\|^2
 =|\braket{x}{\psi}|^2.
\end{equation}
Thus computational-basis measurement probabilities are preserved exactly.  Combined with \cref{cor:circuit}, this proves output-distribution preservation for the circuit language considered here.

\subsection{Locality and overhead}
The source command $A_j$ acts only on party $j$'s data.  The target command \eqref{eq:Ahat} acts only on the same data plus tag $p_j$.  Therefore no command acquires access to another party's register.  The phase code is pre-shared entanglement prepared before the run, so the dynamic party-interaction graph is unchanged.

\begin{theorem}[Ownership and interaction-graph preservation]\label{thm:locality}
For an $n$-party source program consisting of party-local commands and final computational-basis measurements, the compiler adds exactly one two-level real phase tag per party and adds no run-time communication edge.  The target local dimension is doubled at each party, and the compiled program has exactly the same computational-basis output distribution as the source.
\end{theorem}

\begin{remark}[What is known and what is being formalized]
The existence of locality-preserving real simulations is established in \cite{McKague2009}.  \Cref{thm:intertwine,thm:locality} should therefore be read as the semantic correctness theorem of the present \emph{compiler presentation}: it identifies the source command, target command, ownership discipline and explicit phase-code invariant needed to expose that simulation as a misty-language transformation.  We do not claim the underlying real-Hilbert-space embedding as new.
\end{remark}

\section{The GHZ program as an exact gate-level case study}\label{sec:ghz}

We now specialize to three parties, where the target program remains inside the elementary black-and-white gate fragment rather than merely the real-linear intermediate language.

Let
\begin{equation}\label{eq:Gminus}
   \ket{G_-}=\frac{\ket{000}-\ket{111}}{\sqrt2},\qquad
   \ket{G_+}=\frac{\ket{000}+\ket{111}}{\sqrt2}.
\end{equation}
The referee supplies
\[
   q=q_1q_2q_3\in\{000,011,101,110\}.
\]
With the convention of the original misty derivation, the required output parity is odd for $000$ and even for the other three promised inputs.

For $q_j=0$, party $j$ applies $H$ and measures in the computational basis.  For $q_j=1$, the complex source program applies
\[
  S=\begin{pmatrix}1&0\\0&i\end{pmatrix}
\]
then $H$ and measures.  This realizes the $Y$ basis up to an exchange of the two classical labels; promised inputs contain zero or two $Y$ questions, so the label exchange does not change parity.

Since $S_R=\ket0\!\bra0$ and $S_I=\ket1\!\bra1$, \eqref{eq:Ahat} gives the compiled gate
\begin{equation}\label{eq:Shat}
  \widehat S_j
   =\ket0\!\bra0_{d_j}\otimes I_{p_j}
    +\ket1\!\bra1_{d_j}\otimes J_{p_j},
\end{equation}
namely controlled-$J$ from the data token to the local phase tag.

\begin{table}[t]
\centering
\caption{Local translation for the GHZ fragment.  Commands are written in execution order.}
\label{tab:ghzcompile}
\begin{tabularx}{0.94\textwidth}{@{}cXXl@{}}
\toprule
Input & Complex source branch & Real misty target branch & Target owner\\
\midrule
$0$ & $H_{d_j}$; measure $d_j$ & $H_{d_j}$; measure $d_j$ & $d_j$\\
$1$ & $S_{d_j}$; $H_{d_j}$; measure $d_j$ & $\widehat S_j$; $H_{d_j}$; measure $d_j$ & $d_j,p_j$\\
\bottomrule
\end{tabularx}
\end{table}

\begin{proposition}[GHZ sign transfer]\label{prop:sign}
Start the target program in $\ket{G_-}\otimes\ket0_P$.  If no party executes the compiled $Y$ branch, the data state remains $\ket{G_-}$.  If exactly two distinct parties $j,k$ execute it, then
\begin{equation}\label{eq:signtransfer}
  \widehat S_j\widehat S_k
   (\ket{G_-}\otimes\ket0_P)
   =\ket{G_+}\otimes\ket0_P.
\end{equation}
In both cases the phase code finishes unentangled from the data.
\end{proposition}

\begin{proof}
On the $\ket{000}$ branch neither controlled-$J$ fires.  On the $\ket{111}$ branch both fire, and \eqref{eq:twoJ} supplies a factor $-1$ on $\ket0_P$.  That factor cancels the source minus sign:
\[
 \frac{\ket{000}\ket0_P-\ket{111}\ket0_P}{\sqrt2}
 \longmapsto
 \frac{\ket{000}\ket0_P-\ket{111}J_jJ_k\ket0_P}{\sqrt2}
 =\ket{G_+}\ket0_P.
\]
\end{proof}

The final Hadamards expose the parity directly:
\begin{align}
 H^{\otimes3}\ket{G_-}
 &=\frac12(\ket{001}+\ket{010}+\ket{100}+\ket{111}),\label{eq:odd}\\
 H^{\otimes3}\ket{G_+}
 &=\frac12(\ket{000}+\ket{011}+\ket{101}+\ket{110}).\label{eq:even}
\end{align}
The first support is exactly the odd-weight strings; the second is exactly the even-weight strings.

\begin{theorem}[GHZ correctness and locality]\label{thm:ghz}
For every promised input $q\in\{000,011,101,110\}$, the compiled real misty program wins with probability one.  Its output distribution agrees with the complex source program; the phase tags finish in $\ket0_P$ independently of the observed data; and every dynamic operation remains local to one party.
\end{theorem}

\begin{proof}
For $q=000$, no controlled-$J$ is used and \eqref{eq:odd} gives odd parity with probability one.  Each other promised input contains exactly two $1$s; \cref{prop:sign} changes $\ket{G_-}$ to $\ket{G_+}$ and \eqref{eq:even} gives even parity with probability one.  Every allowed output has amplitude $1/2$ and probability $1/4$.  Locality follows from \eqref{eq:Shat} and \cref{tab:ghzcompile}.
\end{proof}

\begin{figure}[t]
\centering
\begin{tikzpicture}[every node/.style={font=\small},
 dbox/.style={draw=black,rounded corners=2pt,fill=white,align=center,minimum width=26mm,minimum height=9mm},
 pbox/.style={draw=figblue,rounded corners=2pt,fill=theoremblue,align=center,minimum width=26mm,minimum height=9mm},
 arr/.style={-{Latex[length=2.4mm]},thick}]
\node[dbox] (a) {$\ket{G_-}\otimes\ket0_P$\\source sign on $\ket{111}$};
\node[pbox,right=7mm of a] (b) {two local controlled-$J$\\$J_jJ_k\ket0_P=-\ket0_P$};
\node[dbox,right=7mm of b] (c) {$\ket{G_+}\otimes\ket0_P$\\phase code factors out};
\node[pbox,right=7mm of c] (d) {$H^{\otimes3}$\\even-parity support};
\draw[arr,draw=figblue] (a)--(b);
\draw[arr,draw=figred] (b)-- node[above=5mm,font=\scriptsize]{sign transfer} (c);
\draw[arr,draw=figgreen] (c)--(d);
\end{tikzpicture}
\caption{The GHZ $Y/Y$ branch.  The 64-path cancellation of the original derivation is compressed to one logical sign-transfer identity.}
\label{fig:ghzflow}
\end{figure}
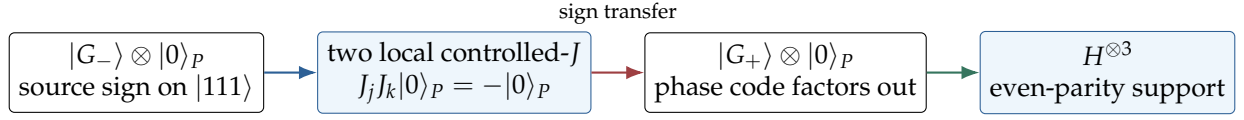

\section{Rewrite-level realization and exact verification}\label{sec:rewrite}

The Hilbert-space proof above is deliberately short, but the construction remains a misty-state rewrite rather than merely a matrix simulation.  This section identifies the exact rewrite obligations and relates them to the longer black-and-white expansion.

\subsection{Local gate identities}
The only nontrivial target operation for the GHZ $Y$ branch is controlled-$J$.  On basis tokens its action is
\[
 \begin{array}{c|cccc}
 (d,p) & (W,W)&(W,B)&(B,W)&(B,B)\\ \hline
 \widehat S(d,p)&(W,W)&(W,B)&(B,B)&-(B,W).
 \end{array}
\]
The control-$W$ cases do nothing; the control-$B$ cases apply $J$ to the phase token.  Because $J=XZ$, each case reduces using only NOT, controlled-NOT, Hadamard-conjugated controlled-$Z$, sign propagation and cancellation.

\subsection{From 64 paths to one invariant}
In the original explicit derivation \cite{ghzbw}, the initial data mist has two components and the phase-code mist has four, producing eight six-token components before local $X/Y$ boxes are expanded.  The Hadamard branches then generate 64 paths.  For the all-$X$ input, terms pair so that even-parity outputs cancel; for an input with two $Y$ questions, the controlled-$J$ action changes the pairing and odd-parity outputs cancel.

The invariant proof performs the same cancellation one level earlier.  Instead of tracking all paths separately, it proves once that the phase-code action on the $\ket{111}$ branch is $-1$ whenever exactly two $Y$ controls fire.  Linearity then gives \cref{prop:sign}, after which the ordinary Hadamard identities \eqref{eq:odd}--\eqref{eq:even} determine the support.  The two proofs are therefore consistent descriptions at different abstraction levels rather than competing derivations.

\subsection{Executable checker boundary}
A small exact checker for this fragment requires no floating point arithmetic.  Terms can be represented as maps from basis strings to coefficients in $\mathbb Q(\sqrt2)$, with rules for tensor product, signed union, $H$, $X$, $Z$, CNOT and cancellation.  The four GHZ inputs then normalize to the supports in \eqref{eq:odd}--\eqref{eq:even}.  Such a checker would mechanize the case-study proof; it is not required for the algebraic correctness theorem and is left as an artifact-level extension rather than being silently assumed here.

\section{Preservation, overhead, and claim boundaries}\label{sec:preserve}

A strong statement of what is preserved is as important as the correctness proof itself.

\subsection{What the compiler preserves}
For the source language of \cref{sec:compiler}, the translation preserves:
\begin{enumerate}[label=(\roman*)]
\item \textbf{denotation under encoding}: $\widehat C\Enc_n(\psi)=\Enc_n(C\psi)$;
\item \textbf{computational-basis probabilities}: by \eqref{eq:prob};
\item \textbf{party ownership}: every command remains owned by the same party;
\item \textbf{dynamic interaction graph}: no new run-time communication edge is introduced;
\item \textbf{asymptotic local overhead}: exactly one two-level phase tag is added per party.
\end{enumerate}
The phase tags may be initially entangled across parties.  This is pre-shared static resource, not run-time communication.

\subsection{What the compiler does not prove}
The construction does \emph{not} prove that two rebits per player are necessary to win GHZ, nor that the chosen phase code is dimension-optimal among all strategies reproducing the same correlation.  Indeed, the triangle graph state
\[
 \ket{G_\triangle}=CZ_{12}CZ_{23}CZ_{13}\ket+^{\otimes3}
\]
is a real three-qubit state.  It is stabilized by $XZZ$, $ZXZ$ and $ZZX$, whose product is $-XXX$; measuring $X$ for one input label and $Z$ for the other reproduces the same odd/even parity pattern.  This is a \emph{strategy realification}.  It does not preserve the specified GHZ source circuit with its $S$ phase operations.

\begin{figure}[t]
\centering
\begin{tikzpicture}[node distance=9mm and 9mm,every node/.style={font=\small},
 box/.style={draw=black,rounded corners=3pt,align=center,minimum width=43mm,minimum height=13mm,inner sep=5pt},
 arr/.style={-{Latex[length=2.4mm]},thick,draw=figblue}]
\node[box,fill=softgray] (goal) {complex GHZ behavior\\and source program};
\node[box,below left=of goal,fill=white] (strat) {strategy realification\\any real realization of\\the same correlations};
\node[box,below right=of goal,fill=theoremblue] (prog) {program realification\\translate the specified state,\\commands and ownership};
\node[below=5mm of strat,align=center,font=\scriptsize] {e.g. three-rebit\\triangle graph state};
\node[below=5mm of prog,align=center,font=\scriptsize] {distributed phase tags\\one local tag per party};
\draw[arr] (goal)--(strat);
\draw[arr] (goal)--(prog);
\end{tikzpicture}
\caption{Two different realification goals.  The six-rebit construction addresses the stronger structure-preserving problem, not minimal dimension for the GHZ correlation.}
\label{fig:two-goals}
\end{figure}
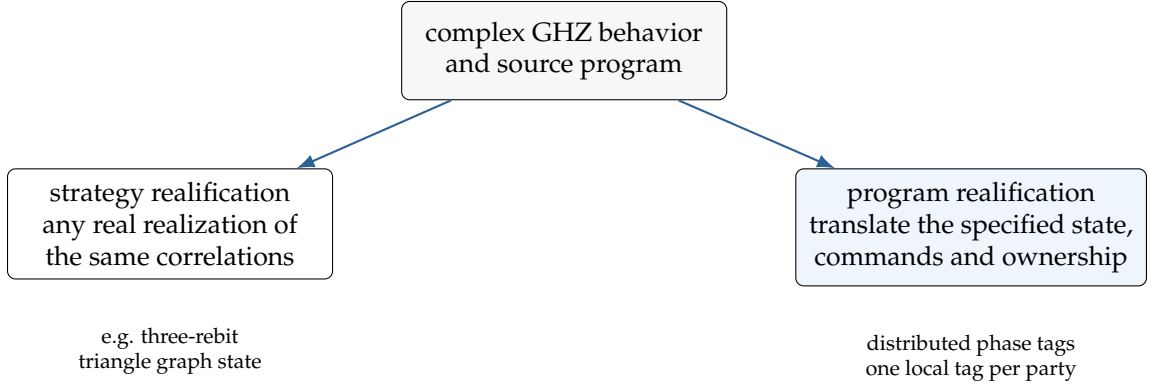

\subsection{Scope of the general compiler theorem}
The proof in \cref{sec:compiler} covers finite pure-state circuits of party-local complex linear maps followed by computational-basis measurements.  Extending the syntax to arbitrary instruments, mixed states, classical control and recursive programs requires the corresponding density-operator or completely-positive semantics.  Nothing in the present theorem establishes such a full language automatically.

Likewise, the semantic real-linear target admits the matrices $A_R$ and $A_I$ as primitives.  The GHZ fragment is stronger in a different sense: its translation is explicitly synthesized in the ordinary black-and-white gate vocabulary.  A general gate-synthesis theorem for arbitrary translated $A$ is a separate compilation problem.

\section{Related work and novelty boundary}\label{sec:related}

\paragraph{Real quantum computation.}
Real-amplitude simulation predates the present misty language.  Rudolph and Grover gave a circuit-model translation using a two-level real/imaginary ancilla and a universal two-rebit gate \cite{RudolphGrover2002}.  McKague, Mosca and Gisin developed a multipartite real simulation in which local evolution and measurements remain local \cite{McKague2009}.  The distributed phase code used here is an explicit presentation of that logical complex structure.  We therefore do not claim that representing complex quantum mechanics over a real Hilbert space, or distributing the complex structure across parties, is new.

\paragraph{Misty-state semantics.}
Rudolph's notation was introduced as an elementary black-and-white route to quantum reasoning \cite{Rudolph2017}.  German's term-rewriting work treats mist expressions as a formal semantic object rather than merely a picture \cite{German2026}.  Newton--German give an explicit amplitude semantics, normalization discipline and exact finite equality machinery \cite{NewtonGerman2026}.  The present work uses that semantic separation to formulate the distributed real simulation as a source-to-target program transformation.

\paragraph{Quantum programming languages.}
Denotational semantics for quantum programming has a long history; Selinger's language and superoperator semantics are a foundational example \cite{Selinger2004}.  Our language fragment is intentionally much smaller.  The goal is not to compete with full quantum languages or graphical calculi, but to make one concrete compilation issue---distributed complex phase in a real rewriting syntax---fully inspectable.

\paragraph{Claimed contribution.}
The result to attribute to this paper is therefore the following package: an explicit $n$-party phase-code invariant in misty notation; a syntax-directed party-local compiler rule $A_R+iA_I\mapsto A_R\otimes I+A_I\otimes J_j$; a direct denotational, measurement and ownership proof for the stated finite circuit language; and a gate-level GHZ realization whose correctness collapses to the two-tag sign lemma.  The paper's contribution is the \emph{program-semantics organization and proof interface}, built on known real-Hilbert-space simulation.

\section{Conclusions and next formalization steps}\label{sec:conclusion}

The black-and-white GHZ construction is most informative when read as a compiler rather than as an alternative hand calculation of a known nonlocal game.  The phase-reference state supplies one logical complex structure through several physically local real actions.  Once that invariant is isolated, the six-token calculation becomes a special case of a simple rule: replace each source occurrence of $i$ in a party-owned operation by that party's local $J=XZ$ action on a distributed phase tag.

For the finite pure-state language considered here, this translation is compositional, probability preserving and locality preserving.  The GHZ case adds an unusually transparent gate-level witness: two local $J$ actions return the phase code to itself with a minus sign, which flips the relative GHZ branch phase and determines the parity after the final Hadamards.  The proof therefore explains the detailed cancellation pattern without requiring the cancellation table to carry the logical burden.

Three extensions now have clear, separable proof obligations.  First, a typed density-operator language could extend the compiler to general measurements, classical feed-forward and mixed states.  Second, a verified rewrite engine could mechanize normalization of the compiled mist terms and serve as an executable certificate layer.  Third, one can ask a genuine minimality question: under explicit requirements of compositionality, party-decomposition preservation and uniform compilation of arbitrary local complex operators, is one local two-dimensional phase system per party necessary?  That is a compiler lower-bound problem and should not be conflated with the much weaker question of how many rebits are needed to reproduce one correlation.

The main conceptual conclusion is correspondingly precise: complex phase is not removed by the real misty semantics; it is \emph{reified as distributed local program state}.  The resulting code makes the scalar action of $i$ locally available to every party without introducing run-time communication, turning a pedagogical black-and-white calculation into a small, auditable example of locality-preserving quantum compilation.

\section*{Acknowledgments} The authors thank Prof. Terry Rudolph for several illuminating conversations that helped crystallize and refine some of the ideas developed in this paper.

\section*{Appendix A. Explicit three-party phase-code check}
For completeness, start from
\[
 \ket0_P=\tfrac12(\ket{000}-\ket{011}-\ket{101}-\ket{110}).
\]
Applying $J_1$ gives
\[
 \tfrac12(\ket{100}-\ket{111}+\ket{001}+\ket{010})
 =\tfrac12(\ket{001}+\ket{010}+\ket{100}-\ket{111})
\]
after accounting for $J\ket1=-\ket0$ on the $110$ term; this is $\ket1_P$.  The same result follows for $J_2$ and $J_3$ by permutation symmetry.  Applying any second $J_k$ with $k\ne j$ gives $-\ket0_P$.  This is the compact algebraic identity that the original four-component phase mist verified term by term.

\section*{Appendix B. Relation to the original path expansion}
The original six-ball derivation \cite{ghzbw} is retained as an independent low-level validation.  Before the final Hadamards, the eight products of the two GHZ data components with the four phase-code components evolve through the local boxes.  For $000$, expansion under three Hadamards yields 64 paths whose even-parity contributions cancel in pairs.  For $011$, $101$ or $110$, exactly two controlled-$J$ operations alter the relative sign on the $\ket{111}$ data component; the pairing reverses and odd-parity contributions cancel.  \Cref{prop:sign} factors this calculation through the logical code space: the 64 individual paths are the distributive expansion of the single identity $J_jJ_k\ket0_P=-\ket0_P$ followed by \eqref{eq:odd} or \eqref{eq:even}.

\vspace{0.6em}
\hrule
\vspace{0.5em}
{\footnotesize
All computational algorithms, proof-search procedures, and certificate criteria were developed by the author(s). Large Language Models (LLMs) such as ChatGPT and Claude were utilized strictly as computational assistants for debugging code and cross-checking routines. No mathematical claims rely on LLM outputs; all results are grounded exclusively in independently reproducible, rigorously verified certificates (exact, symbolic, rational, interval, exhaustive, or controlled high-precision).
}

\end{document}